\documentclass[12pt, letterpaper]{article}
\usepackage{graphicx}
\usepackage{color}
\usepackage{wrapfig}
\usepackage{hyperref}

\begin{document}

\title{
		Universal and specific features of Ukrainian economic research: publication analysis based on Crossref data%
		%
		\thanks{This work was supported in part by the National Research Foundation of Ukraine, project No.~2020.01/0338.}
	}
	

	\author{O.~Mryglod$^1$ \and S.~Nazarovets$^2$ \and S. Kozmenko$^3$ }
	
	\date{February 24, 2021}

	\maketitle
	
	$^1$ Institute for Condensed Matter Physics of the National Academy of Sciences of Ukraine, 1 Svientsitskii St., 79011 Lviv, Ukraine
	
	$^2$ State Scientific and Technical Library of Ukraine, 
	180 Antonovycha Str., 03680 Kyiv, Ukraine
	
	$^3$ University of Social Sciences Spoleczna Akademia Nauk, 9 Sienkiewicza St., 90--113 {\L}\'{o}d\'{z}, Poland\\
	
		\textit{This is the version of the Article before peer-review has taken place. The paper is submitted to \textit{Scientometrics} journal}
		
	\begin{abstract}
		Our study is one of the first examples of multidimensional and longitudinal disciplinary analysis at the national level based on Crossref data. We present a large-scale quantitative analysis of Ukrainian economics. This study is not yet another example of research aimed at ranking of local journals, authors or institutions, but rather exploring general tendencies that can be compared to other countries or regions. We study different aspects of Ukrainian economics output.  In particular, the collaborative nature, geographic landscape and some peculiarities  of  citation statistics are investigated. We have found that Ukrainian economics is characterized by a comparably small share of co-authored publications, however, it demonstrates the tendency towards more collaborative output. Based on our analysis, we discuss specific and universal features of Ukrainian economic research. The importance of supporting various initiatives aimed at enriching open scholarly metadata is considered. A comprehensive and high-quality meta description of publications is probably the shortest path to a better understanding of national trends, especially for non-English speaking countries. The results of our analysis can be used to better understand  Ukrainian economic research and support research policy decisions.
	\end{abstract}
	
	\section{Introduction}
	Research process is a complex system that includes human natural activity, uncertainty of creative process, non-trivial patterns of collaborative structures and many other aspects, which makes any unambiguous formalization impossible. Therefore, both for countries with established economies and ``healthy'' research ecosystems and for developing countries that are just initiating required reforms, there are no universal and simple approaches to research evaluation. 
		Assessment of research outputs, which are mainly reflected in the form of scholarly publications, is a widely accepted way of evaluating the efficiency of the research process itself. However, 
	it is far from being simple to combine expert judgements and quantitative methods. Moreover, any implemented decision causes the immediate and obvious or remote and hidden reaction of the entire system and its further adaptation. Sometimes, the change is rather natural. For example, social sciences and humanities (SSH), where output is traditionally more diverse in terms of language, publication types, etc., demonstrate the tendency to increase the share of journal papers and English-language publications  \cite{Kulczycki2018}. Sometimes, such adaptation is expressed in a completely unexpected way. The haphazard application or even misuse of quantitative metrics can lead to the resistance of researchers and even publishers. The absence of editions in official registers or authoritative abstract databases can be represented even in a positive sense, as an indication of some particularity. The problem is that such a nonconformity of periodicals is usually accompanied by invisibility: ISSN for journals, DOI for papers, ORCID for authors, a web-page with all necessary information available, and other formal elements  make particular edition an element of the entire system. The more data are available, the more comprehensive analysis can be performed in order to understand the typical features of scientific disciplines and the role of the considered journal, to follow the evolution of knowledge structure, to map the national or, more globally, world science. 
	
	Today, we are witnessing the deep crisis of Ukrainian scientific research system: permanent under-funding, outdated management methods, weak integration in the world information space and many other problems form the vicious circle \cite{Schiermeier2019,Hladchenko2020}. We believe that the situation cannot be improved without a clear understanding of the current state of national research. Unfortunately, only the fragments of relevant information can be found from various sources. Some disciplines that are related to ``hard'' sciences are traditionally better ``visible'' as they are  better covered by Scopus and Web of Sciences (WoS). Such disciplines also tend to be more willing to follow new rules of the game, such as maintaining web-pages, publishing preprints, assigning DOI, etc. For example, the analysis of openness of Ukrainian academic journals \cite{Mryglod2012} revealed that the ``most visible'' SSH section is History, Philosophy and Law. But only 25\% of its publications on average were fully available online. The corresponding value for the Physics and Astronomy section was almost 45\%; for the Chemistry section --- almost 35\%, and for Mathematics --- 32\%. The large-scale quantitative analysis of SSH often cannot be done due to the lack of available data. But new possibilities appear with the development of new information tools. The scientific research system is gradually adapting to the challenges of new times.

The purpose of this paper is to provide a quantitative description of the Ukrainian journal papers in the field of Economics using Crossref data.
	Ukrainian economic discipline is chosen as a case study  for many  reasons. Ukrainian research is understudied in general, and there are no examples of a quantitative scientometric study at the  national level so far. The results of such analysis can be used in practice to better understand the selected area of Ukrainian research, develop benchmarks necessary to use metrics and support research policy decisions. Such a case study will make Ukraine more visible on the world map providing the description of another Eastern European developing country with features such as Cyrillic writing and non-English speaking.
	
	Economics is chosen as a representative of SSH, which is considered ``as the bridge between the `hard' sciences and the `social' sciences'' \cite{Cainelli2012}.
	To quote a very nice formulation in \cite{Sasvari19}, this paper is aimed at accumulating the ``evidence-based assessment'' of publishing behavior in Ukrainian economics. 
	
	The paper is organized as follows: some relevant peculiarities of Ukrainian research policy and Ukrainian economic research are discussed in the next section. The data set used in our case study is described in section~\ref{data_sec}. A brief discussion is given in section~\ref{results_sec} along with a comparative analysis performed at the publication level for each of the obtained results. General conclusions are drawn in section~\ref{Conclusions_sec}.
	
	\section{Research in Ukraine. Ukrainian economic research}
	
	Economics is a popular discipline for education and research in Ukraine. It is in the TOP3 list of disciplines (the only from SSH range) by the number of doctoral theses defended during 2018--2020 in Ukraine according to the statistics of the Ministry of Education and Science of Ukraine \cite{RadeykoBlog}. This automatically means that a large number of journal papers are published: the necessary minimum to get a doctoral degree is officially determined \cite{Order2019}. Publications also form the basis for many other aspects of research evaluation. Therefore, the publishing behaviour is both consciously and unconsciously  governed by the external rules. To take this into account for  further interpretations, the main peculiarities of the latter have to be discussed. 
	
	Over 2.5 thousand of titles are listed on the web portal ``Scientific Periodicals of Ukraine'' managed by Vernadsky National Library of Ukraine. Most of these journals publish papers exclusively in Ukrainian; they are not indexed in important international abstracting and indexing databases. Therefore, one can assume that these numerous journals  disseminate scientific information mainly at the regional level, reaching only limited audience \cite{Sivertsen16} and performing their special functions \cite{Moed2021}. 1,276 editions are officially recognized (the most covered SSH disciplines in this context are Economics with 227 relevant journals and Law with 98)\footnote{in January 2021}: the official List of scientific professional editions of Ukraine is regularly updated. Only papers in these Ukrainian journals are taken into account in many assessment procedures. The rules for recognizing publications in foreign sources were different in different time periods.  
	
	It is interesting to note that due to illiterately formulated requirements, all publications in foreign journals were automatically treated as appropriate independently on their quality during a long time. Moreover, the analysis of the Ukrainian policies  between 2012 and 2018 (e.g., \cite{Order2012,Order2018}) reveals the implicit stimulus to formal publishing in foreign journals, but no criteria to select such journals were provided. Therefore, one can speculate about numerous papers in low-quality and ``invisible'' journals published abroad. Since then, many changes were adapted to encourage publishing in journals indexed in Scopus and WoS databases \cite{Order2012}. However, the share of papers by Ukrainian authors indexed in these databases is still small. Interestingly, the majority of such publications can be found again in national journals \cite{Nazarovets2019}, in spite of the bias towards these editions (see also  \cite{Sasvari19}). 
	To give an example, according to WoS data, over 60\% of all economic papers where Ukraine is mentioned in the affiliations are published in Ukrainian journals; and this share is greater in Scopus (70\%)\footnote{data assessed in January 2021}.

	\section{The Data}\label{data_sec}
	
	Before describing the data set in our case study, some  \emph{limitations} have to be mentioned. 
	(i) First of all, only journal papers are taken into account. While presenting results in a form of books is significant for Economics, journal papers can still be considered as the dominant type of output. It is found that two-thirds of publications in the social sciences consist of journal papers \cite{Sivertsen16}. Moreover, book metadata are hardly available and, therefore, cannot currently be used for analysis. Ukrainian information services provide only limited portions of bibliographic information: non-structured and published with large time delay, it can be used only to draw general retrospectives. Book metadata 
	can be found in the Ukrainian abstract database \emph{``Ukrainika Naukova''} \cite{Ukrainika}. However, no export of structured metadata is available, and the database is out of date.
	Also, the annual list of all journal articles and  monographs within different research disciplines are prepared by the \emph{``Ukrainian book chamber''} service \cite{BookChamber}: the so-called \emph{litopysy} (chronicles) contain bibliographic records. However, digital versions of these unstructured lists are published once a year with a two-year delay. According to the data on Ukrainian economic research retrieved from these two sources,  statistics of journal papers are about twice as large. 
	
	(ii) Only \emph{papers in the journals published in  Ukraine} are included in the main data set. Theoretically, to cover all outputs of Ukrainian economic research, it is necessary to consider papers by Ukrainian authors in both national and foreign editions. But althought there is a possibility to cover many of Ukrainian journals (see below), it is hard to collect necessary metadata for the papers published in journals of other countries, which are outside the internationally recognized abstract data bases. However, Scopus data related to economic publications by Ukrainian authors in both national and foreign journals are partially used for particular tasks of research. 
	
	The \emph{Crossref} (https://crossref.org) is used as a source of scholarly metadata in this study. Starting from 2018 \cite{Order2018}, the assignment of DOI to each publication is one of the requirements for Ukrainian journals to be included into the national List. This means that relevant structured metadata are delivered to Crossref systematically. Moreover, many leading Ukrainian publishers actively support the Initiative for Open Citations (I4OC; https://i4oc.org), and more than 500 million references are now openly available through the Crossref API \cite{Peroni2020}. Although Crossref avoids creating any metrics or special analytics, other toolmakers actively use this open citation metadata in their services (e.g., Dimensions, Lens.org, or COCI)  \cite{Hendricks2020}. 
	The developers of the Open Ukrainian Citation Index (OUCI; https://ouci.dntb.gov.ua/en/), a search engine and a citation database that uses open Crossref data, also followed this path.
	The OUCI interface includes search filters, special for Ukrainian publication data. Moreover, additional journal information is taken into account: e.g., categories defined by Ukrainian policy or specialties according to the classification recognized by the State Attestation Commission of Ukraine. In addition to the mission to provide simple and contextual search of Ukrainian scholarly data, the OUCI is aimed at attracting publishers' attention to the problem of completeness and openness of the scholarly metadata \cite{Cheberkus2019}. Crossref classifies documents into broad disciplines \cite{Visser2021}, but most documents of our data set were not marked according to this classification. Thus, the OUCI provides an unprecedented opportunity for Ukrainian data (only at the journal level).

\begin{figure}[ht]
		\vspace{-6mm}
		\centerline{\includegraphics[width=5cm]{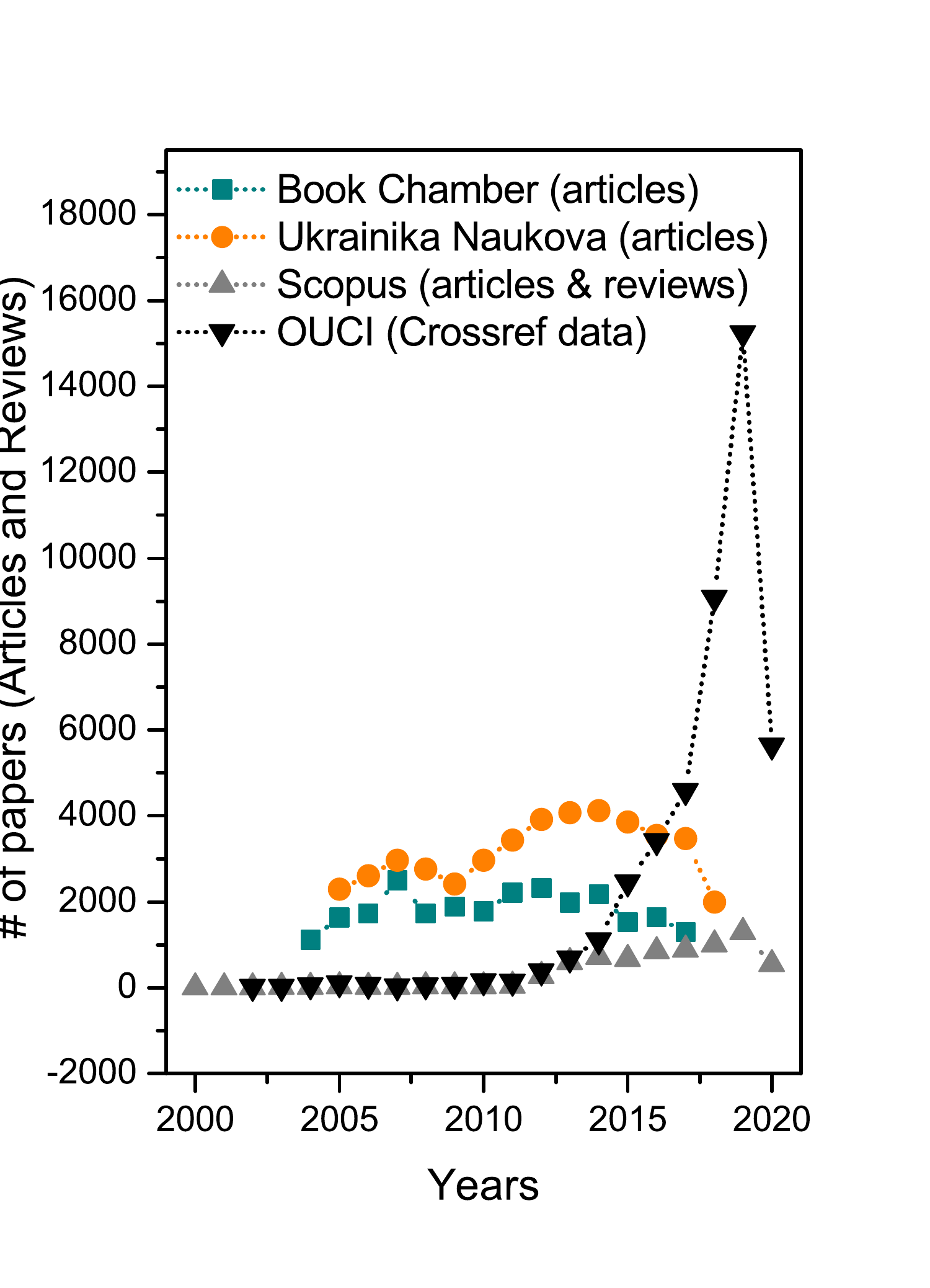}}
		\caption{The annual numbers of paper records related to Ukrainian Economics retrieved from different sources (August 2020): Book Chamber service (disciplinary catalogs are available); Ukrainika Naukova database (internal topical classification indices assigned to papers are used for search); Scopus database (Search phrase: ``AFFILCOUNTRY ( UKRAINE )  AND  ( LIMIT-TO ( SUBJAREA ,  ``BUSI'' )  OR  LIMIT-TO ( SUBJAREA ,  ``ECON'' ) )''; OUCI (journal specialties include ``Economics (code 051)''). }
		\label{Fig_annual_diff_srcs}
	\end{figure}

	According to OUCI, over 300 thousand of papers in over 1,550 Ukrainian editions (367 publishers) are indexed in Crossref\footnote{February 2021}. And this statistics is exponentially increasing. Therefore, OUCI interface allows one to collect comparably large statistics on journal publication metadata related to Ukrainian publishers. The usefulness of OUCI is demonstrated in Figure~\ref{Fig_annual_diff_srcs}: the annual numbers of economic journal papers by Ukrainian authors obtained from different data sources can be compared. It is worth noting that disciplinary labeling is performed differently in each case --- this is one of the possible reasons for different statistics. A journal-level classification is used in the case of OUCI: specialities are manually defined by OUCI staff following the description of the ``Aims and Scope'' section on journals’ web-pages. The obvious drawback here is the inability to separate purely economic papers published in multidisciplinary journals.

	The State Attestation Commission of Ukraine defines Economics discipline through a number of specialties. Accordingly, the OUCI search was narrowed to the following ones: Economics; Tax and Accounting Policy; Finance, Banking and Insurance; Management; Marketing; Business, Entrepreneurship and Stock Markets; Public Administration; and International Economic Relations. In addition, only journals from the National List are considered. 56,301 records were retrieved in December 2020. However, some journals are attributed to many specialties being multidisciplinary. Preliminary analysis and manual inspection of initial data set showed that the topics of many papers are too far from Economics. To end up with the most relevant and representative journals, additional filtering was applied. One of the following conditions is expected to be satisfied: (i) journal specialties are limited by the list mentioned above; (ii) only four journal disciplines (upper classification level) --- ``Social and Behavioral Sciences'', ``Management and Administration'', ``Public Management and Administration'' and ``International Relations'' --- are exclusively taken into account. The final data set includes 23,964 publication records related to 123 Ukrainian journals. Publications period is limited to 2002--2020 years.
	
	\section{Results} \label{results_sec}
	
	\emph{Publication statistics.} The annual number of publications in our data set rapidly increases after 2012: while only 69 papers per year on average are found for the period 2002--2012, the same value for 2013--2020 already exceeds 2,900. Therefore, the latter data subset is considered in the cases when annual effects are investigated.

\begin{figure}[ht]
	\centerline{\includegraphics[width=5cm]{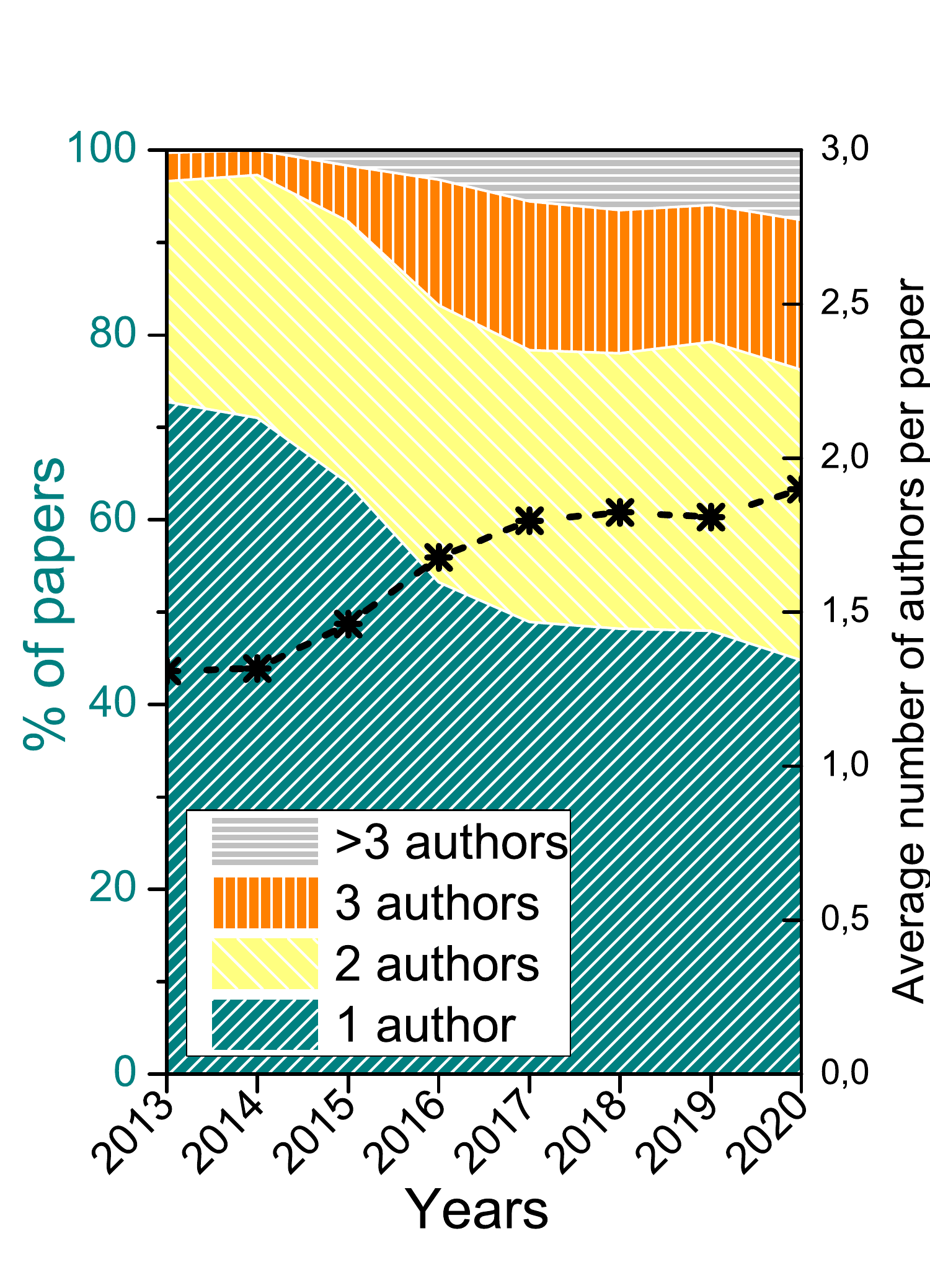}}
	\caption{The annual shares of Ukrainian economic papers categorized by the sizes of co-authorship teams, i.e. consisting of 1, 2, 3 or at least 4 authors. The corresponding average numbers of authors per paper are indicated by black symbols.}
	\label{Fig_Coauth_per_Pap}
\end{figure}

	\emph{Size of co-authorship teams.} 
	A typical Ukrainian economic paper is written by one (almost 50\% of publications) or two (30\%) authors. The largest collaboration team consists of 12 authors. The well-known tendency towards more collaborative economic research (e.g., see \cite{McDowell1983,Hudson1996,Nowell2011,Laband2000,Kuld2018}) can be observed for Ukrainian data. Following the definitions in \cite{Laband2000}, an increase in the incidence of co-authorship and in the extent of co-authorship can be observed (see Figure~\ref{Fig_Coauth_per_Pap}). Not only the noticeable increase of annual \emph{share of collaborative papers} is observed since 2014, but the sizes of co-authorship teams also increase. Moreover, some effects are even more remarkable for Ukrainian data compared to others. To give an example, doubling of shares of publications by three, four or more authors during $\approx 15$ years (1996--2015) is stated in \cite{Kuld2018}, but this share for Ukrainian data increased from  $16.8\%$ to $23.8\%$ (over 1.4 times greater) during only 5 years (2016--2020).

\begin{figure}[ht]
		\includegraphics[width=0.45\textwidth]{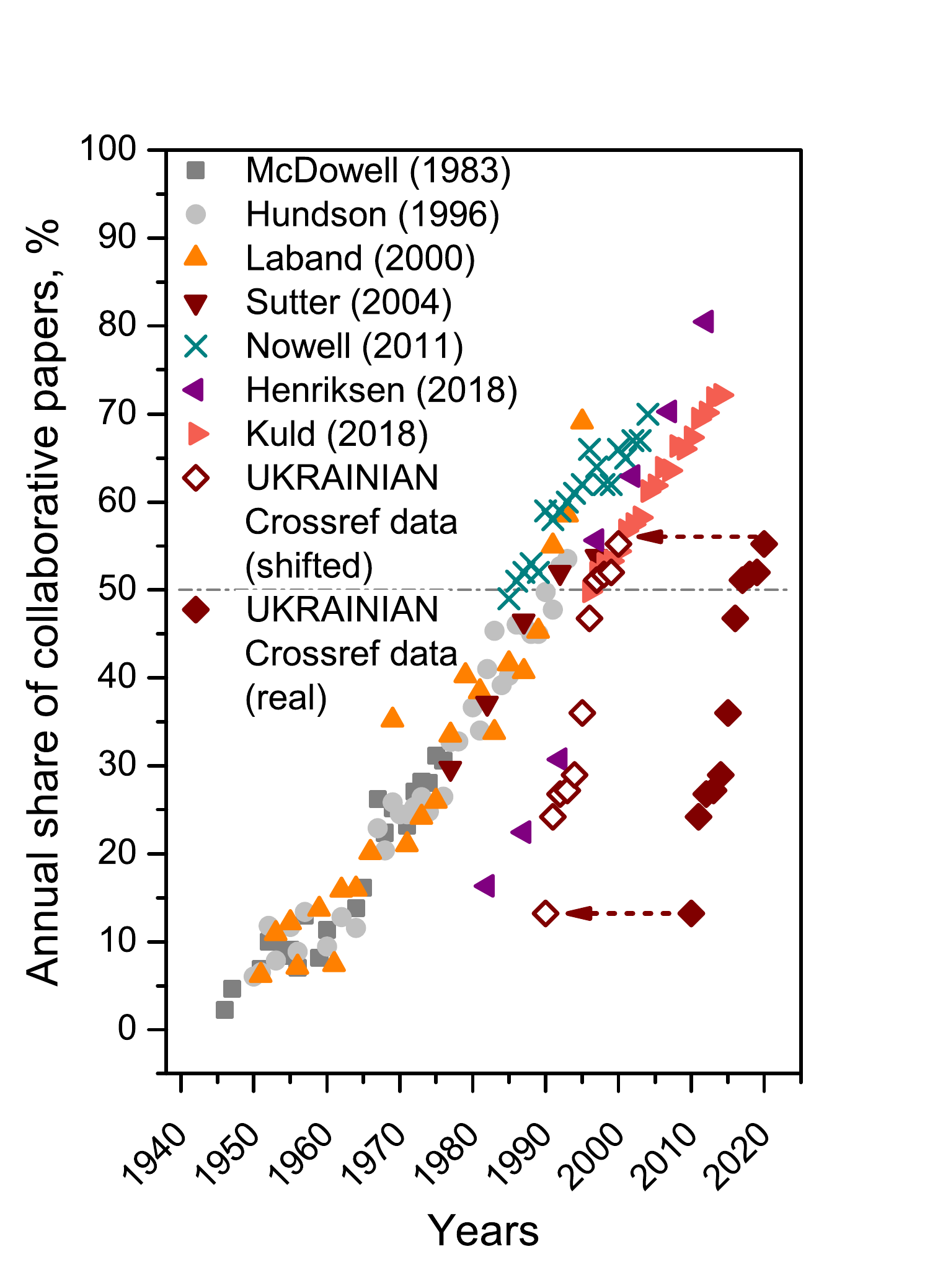}%
		\hfill%
		\includegraphics[width=0.45\textwidth]{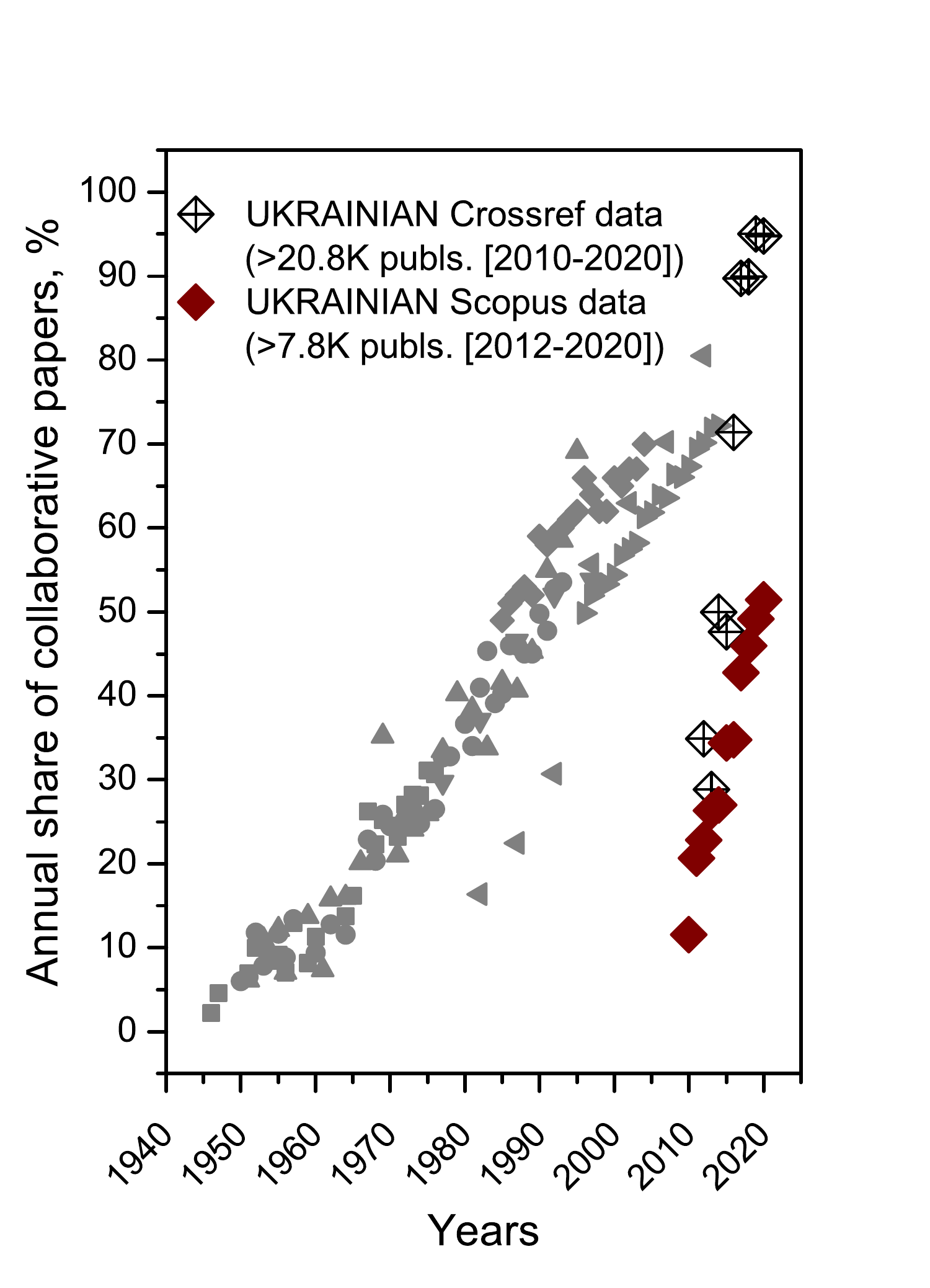}%
		\\%
		\parbox[t]{0.48\textwidth}{%
			\centerline{(a)}%
		}%
		\hfill%
		\parbox[t]{0.48\textwidth}{%
			\centerline{(b)}%
		}%
		\caption{The annual shares of co-authored economic publications for different data sets: geometric symbols [the shapes are the same for (a) and (b)] denote data points taken from texts or digitized plots (WebPlotDigitizer is used, \cite{WebPlotDigitizer}) in \cite{McDowell1983,Hudson1996,Laband2000,Sutter2004,Nowell2011,Henriksen2018,Kuld2018}. The data for Ukrainian economic research are shown by diamons: (a) solid for the entire set of real Crossref data and open for manually shifted ones; (b) solid for real Crossref data excluding the publications indexed also in Scopus or WoS and crossed for Scopus data.}
		\label{Fig_old-our_comparison}
	\end{figure}

	However, while general trends are similar, the particular characteristics of Ukrainian data can be found. For example, while 70--80\% of collaborative papers in Economics in 2010--2014 were reported \cite{Kuld2018,Henriksen2018} for other data sets, the highest mark reached for Ukrainian data during 2017--2020 is about 
	52.5\%. According to other results, late 1990s is the latest period when the share of economic solo-publications was found to be at least 50\% (see Figure~\ref{Fig_old-our_comparison}~(a)).  The typical trend is partially reproduced  for Ukrainian economic research only if its data are manually  shifted back along the timeline. Similar situation is found for the \emph{average number of co-authors per one publication}. The gradual growth of this value is observed: it was close to 1 before 1950, 1.5--1.7 in 1990s \cite{Laband2000,Nowell2011,Kumar2014}, and 1.9 in 2000s  \cite{Nowell2011,Henriksen2018,Kumar2014}. After 2010, economic publications are found to be written by two authors on average \cite{Henriksen2018,Rath2016}. The corresponding value for Ukrainian data is almost 1.8, which again is typical for late 1990s. 
	
	Different data sources and, thus, different disciplinary classifications are used to obtain trends depicted in Figure~\ref{Fig_old-our_comparison}~(a). Previously published results concern the data related to top journals and, presumably, to output coming from ``advanced'' countries (the dominant contribution is typically made by the authors from the US, see \cite{Sutter2004}). On the contrary, almost 87\% of publications in our data set are not indexed in Scopus or WoS databases. Another difference is that other studies \cite{McDowell1983,Hudson1996,Nowell2011,Laband2000,Henriksen2018,Kuld2018,Sutter2004} are often centered on the top economic journals rather than on the data related to a particular country. The exception is \cite{Henriksen2018}, where the search was performed by authors' affiliation country (Denmark) --- the corresponding imaginary slope for resulting data is the most similar to the one for Ukrainian data (Figure~\ref{Fig_old-our_comparison}~(a)). 
	
	It is curious to speculate about the reasons for such ``delayed'' trends observed for Ukrainian economic research. The underlying mechanisms that govern co-authorship are discussed in many papers. There is no need to repeat everything here, we will only mention just some of them \cite{Hudson1996,Laband2000,Nowell2011,Henriksen2016}. Firstly, an increase in specialization and complexity of the Economics discipline itself implies the need to combine professional skills of scholars. Secondly, the increased use of quantitative approaches makes it easier to divide work tasks. In addition, ``the great overlap between quantitative methods and empirical articles'' is declared in \cite{Henriksen2018}. Positive correlations between international collaboration and the probability to get published in top journals are also discussed \cite{Henriksen2018,Kuld2018}. Therefore, one can suggest the dominance of theoretical studies and the prevalence of qualitative approaches in Ukrainian economic research. However, it is not obvious whether such a conclusion can be drawn for the entire Ukrainian economic research or only for  part of it (large in our case). To our knowledge, there is a lack of analysis of economic publications not indexed in Scopus or WoS for other countries \cite{Kulczycki2018}. 
	
	To be able to use the Ukrainian data for more accurate comparison, additional search for internationally visible publication data was performed. 8,215 records are collected from Scopus in January 2021 using the following search request: (SUBJAREA (BUSI) OR SUBJAREA (ECON) AFFILCOUNTRY (UKRAINE)) AND (LIMIT-TO (SRCTYPE, ``J'')). It can be seen in Figure~\ref{Fig_old-our_comparison}~(b) (publications indexed in Scopus or WoS are excluded from the Crossref data set to make this particular subfigure) that although Ukrainian Economics became visible in Scopus later, it demonstrates surprisingly rapid growth in collaborativeness, approaching 95\%. 
	
	Thus, two current characteristic ratios of collaborative publications are observed for Ukrainian economic research. The higher share of co-authored papers in the journals indexed in Scopus corresponds to current world trends. The lower share of co-authored outputs is found for journals invisible in Scopus and WoS databases. 
	One can speculate about the specialization and disciplinary fragmentation of economic research \cite{Henriksen2018,Fourcade2015} supposing that some of topics are more locally nested and, therefore, are targeted at particular audience, while others are more requested on the international level. 
	For example,  such a diversification of topics depending on journals' venues is mentioned in \cite{Springer2019} (p.~16).
	If this is the case for Ukraine, one can also suppose that the topics of international interest are more related to quantitative methods and experimental research. 
	Having no comprehensive evidence for different countries, it is hard to say if such situation is typical for economic research in general. It is also not clear what proportion of papers in the local vs. internationally recognized journals is typical for economic research in other countries. There is some evidence that a relatively small share of SSH publications from national databases are represented in WoS: e.g., 15\% for Poland and 25.9\% for Slovakia \cite{Kulczycki2018}. In our data set, only 5.6\% of publications are indexed in the WoS, but our dataset contains only journals of Ukrainian publishers. The Ukrainian Scopus statistics used for Figure~\ref{Fig_old-our_comparison}~(b) is almost three times smaller compared to the Crossref statistics related to local journals.

In addition, different authors' motivations can be hidden behind the increase in co-authorship. It was already mentioned that external factors can change the natural publishing behavior of researchers. Currently, publications in the internationally  recognized journals (i.e., indexed in Scopus or WoS) are prioritized by Ukrainian Policies. Presumably, it is easier to perform the high-quality research and, therefore, to get published in top journals for a team rather than for an individual. Moreover, one can speculate about financial benefits of collective submission. Due to the persistent under-funding of Ukrainian research, it is hardly possible to cover article processing charges (if any), especially in foreign journals, using institutional funds. Another interesting question is to check, if the publishing behaviour of authors differs concerning OA-journals, which often use the ``Golden'' principle of charging authors for manuscript processing. 
\begin{figure}[h]
	\centerline{\includegraphics[width=5cm]{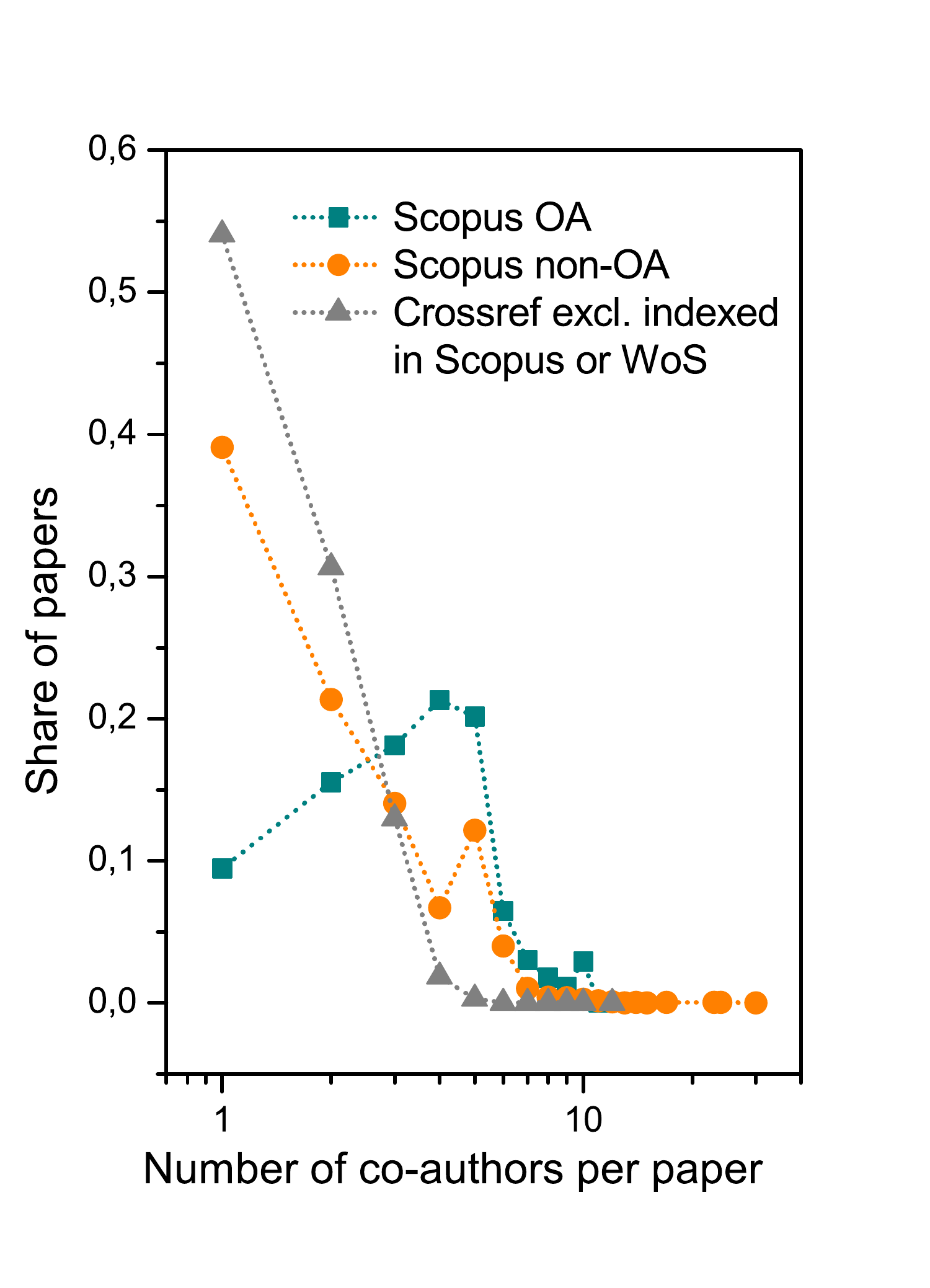}}
	\caption{The distributions of Ukrainian economic publications by the number of co-authors based on Crossref data (excluding publications which are also indexed in Scopus or WoS) --- squares; Scopus data only for OA-journals --- circles; Scopus data only for non-OA-journals --- triangles.}
	\label{Fig_APC}
\end{figure}
To find an answer, the distributions of papers by the number of co-authors for different data subsets are compared in Figure~\ref{Fig_APC}. One can see that while the shapes are more similar for Crossref and Scopus non-OA publications, a different one is observed for Scopus OA data. The typical values can be derived for the latter: more than  40\% of all OA papers are written by groups of four or five authors. On the other hand, the distribution for OA papers is characterized by short tail indicating a natural upper limit. Presumably, too large co-authorship team makes individual benefit from OA paper insignificant. It is also  harder to manage fair sharing of expenses within the large group of contributors. Nevertheless, there is a reason to accept the hypothesis about a  significant role of economic benefit from collective submission of OA papers.

\begin{figure}[ht]
		\centerline{\includegraphics[width=0.6\textwidth]{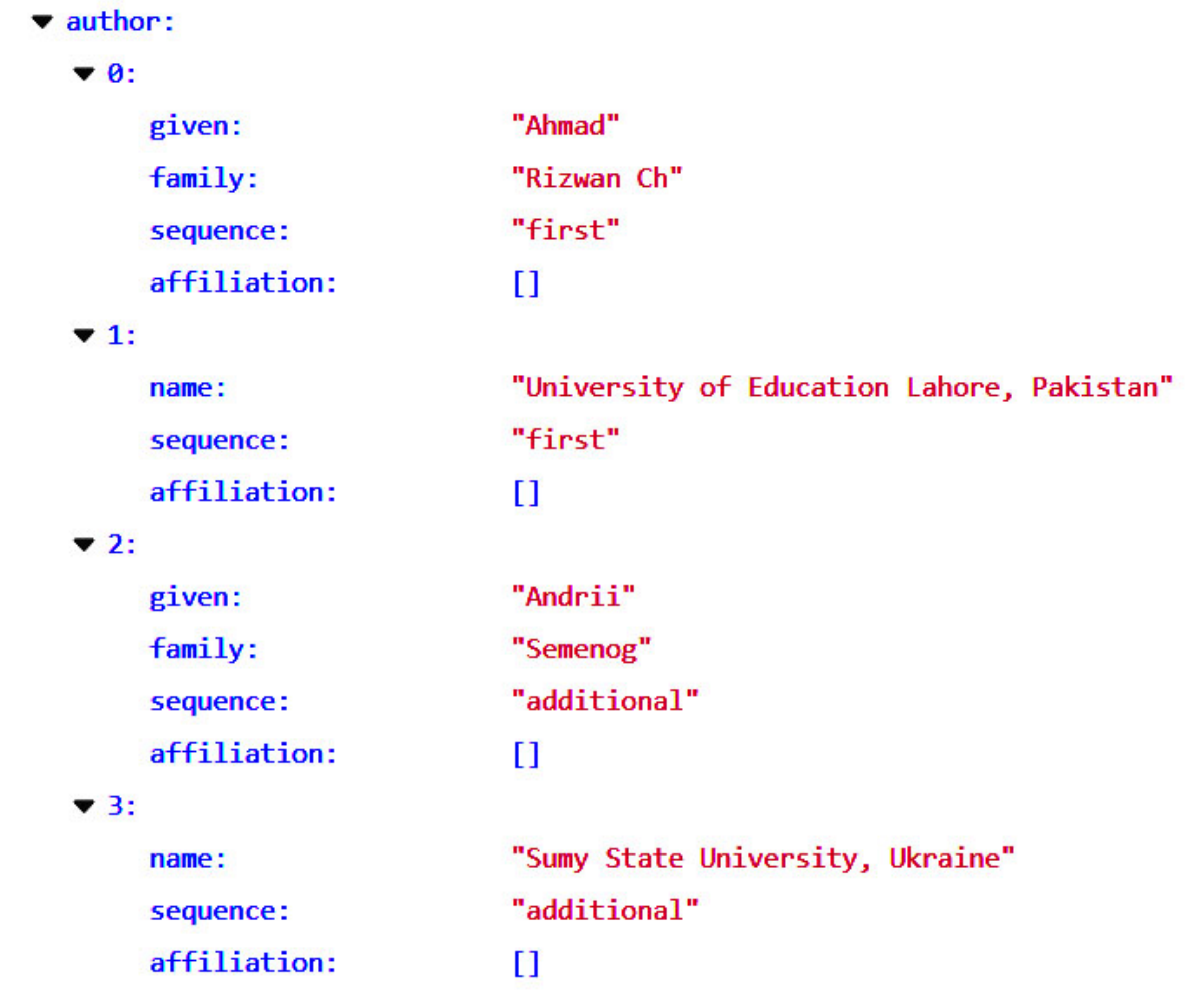}}
		\caption{An example of incorrectly deposited Crossref metadata: two authors and their affiliation data are described as four separate authors.}
		\label{Fig_incorrect_affil}
	\end{figure}

	\emph{Geographic landscape.} Since affiliation field is not included in OUCI export, Crossref API was used to acquire these data for the publications in the data set. Moreover, affiliation data can be partially reconstructed from authors' fields due to incorrectly deposited metadata (see Figure~\ref{Fig_incorrect_affil}).
	Affiliation information was obtained for $\approx 58.4\%$ of records in this way. Using a prepared list of keywords, contributing countries were detected for 13,985 papers. Within this subset, the majority of papers are related to Ukraine: almost 82\% of all contributions are made by Ukrainian authors only, and almost 2\% are submitted in international collaboration. %
	In general, the geographical spectrum is diverse. 
	While  data statistics is very poor until 2012--2013, already 12 foreign countries that made contributions to Ukrainian economic research were found for this period. Presumably due to the more active depositing of Crossref metadata, the explosion of the general number of countries is observed in 2016: it tripled starting from 22 and ending with 65. This number is increasing further until 107 in 2020 (on average, authors from 75 countries contribute annually). However,  only few foreign countries provided at least 1\% of all publications in Ukrainian economic journals. TOP10 of foreign countries by the number of contributions is as follows: 
	\begin{itemize}
		\item SOUTH AFRICA	324	(2.3\%) publications -- 7 of them are co-contributed with Ukraine;
		\item POLAND	193	(1.4\%) -- 91;
		\item INDONESIA	158	(1.13\%) -- 2;
		\item USA	143	(1.02\%) -- 20;
		\item NIGERIA	109	(0.78\%) -- 4;
		\item SLOVAKIA	107	(0.77\%) -- 10;
		\item INDIA	102	(0.73\%) -- 1;
		\item RUSSIA	94	(0.7\%) -- 13;
		\item CZECH REPUBLIC	79	(0.56\%) -- 14;
		\item JORDAN	70	(0.5\%) -- 2.
	\end{itemize}

	Given that some papers can be attributed to more than one Part of the World, it is found that foreign European and Asian countries have contributed equally (38.6\% of papers are related to each category). Almost one fifth (19.8\%) of publications correspond to African countries (the highest peak is found in 2016), over 6.8\% -- to North America and small shares to Oceania (1.4\%) and South America (0.94\%).

	The annual share of internationally collaborative papers (two or more countries are involved) is small: less than 5.5\% since 2016 and even lesser before. In general, 570 (4\%) of papers are characterized by bilateral international collaborations and 54 ($0.39\%$) are related to three countries. The largest number of countries per one paper is four: two such publications are found (BAHRAIN--EGYPT--TURKEY--LITHUANIA and CZECH REPUBLIC--MOLDOVA--UKRAINE--POLAND).

	\begin{table}[htb]
		\caption{The statistics of international contributions within Ukrainian economic publication data sets: (i) all the papers  labelled by countries; (ii) only papers that are also indexed in Scopus or WoS; (iii) only papers NOT indexed in Scopus or WoS.}\label{Tab_Dbs_nonDbS}
		\begin{tabular}{|p{4cm}|p{2cm}|p{2cm}|p{2cm}|}
			\hline Category & \multicolumn{3}{|c|}{Number (\%) of papers within subsets}\\
			\hline
			& All & Indexed in Scopus or WoS & Not indexed in Scopus or WoS  \\
			\hline
			Single Ukrainian author& 11430 (81.7\%) & 529 (24.3\%) & 10901 (92.3\%)\\
			\hline
			Single foreign author& 1929 (13.8\%) & 1264 (58.2\%) &  665 (5.6\%)\\
			\hline
			Two or more authors, including from Ukraine& 346 (2.5\%) &159 (7.3\%) & 187 (1.6\%)\\    
			\hline
			Two or more foreign authors& 280 (2\%)& 221 (10.2\%)& 59 (0.5\%)\\
			\hline    
		\end{tabular}
	\end{table}
	It is expected that the share of papers written in the international collaboration is higher for publications in high-level journals, in particular in those indexed in Scopus and Wos (international co-authorship is even encouraged by these services). To give an example, over half of economic publications from ASEAN countries (1979–2010) are internationally collaborative in \cite{Kumar2014}. To some extent, this expectation is also satisfied  for Ukrainian data (see Table~\ref{Tab_Dbs_nonDbS}). Having in mind that many of countries do not have common papers with Ukraine, one can speculate about different motivations of foreign contributors: while some of them are interested in collaboration with Ukraine, others can consider Ukrainian journals (especially those indexed in the international databases) as an additional platform for publishing own results. Some hints can be derived by comparing the lists of foreign countries that are top contributors or/and  top collaborators. One third of foreign countries are not connected by collaborative papers with Ukraine.

	To investigate the structure and the strength of international collaboration links in Ukrainian economic journals, the co-authorship network at the level of countries was built. Each of 107 nodes represents a particular country that has contributed at least once. The link connects two countries if they are found in one paper at least once. Link width (strength) is proportional to the number of common publications. 
	
	\begin{wrapfigure}{o}{5cm}
		\vspace{-6mm}
		\centerline{\includegraphics[width=5cm]{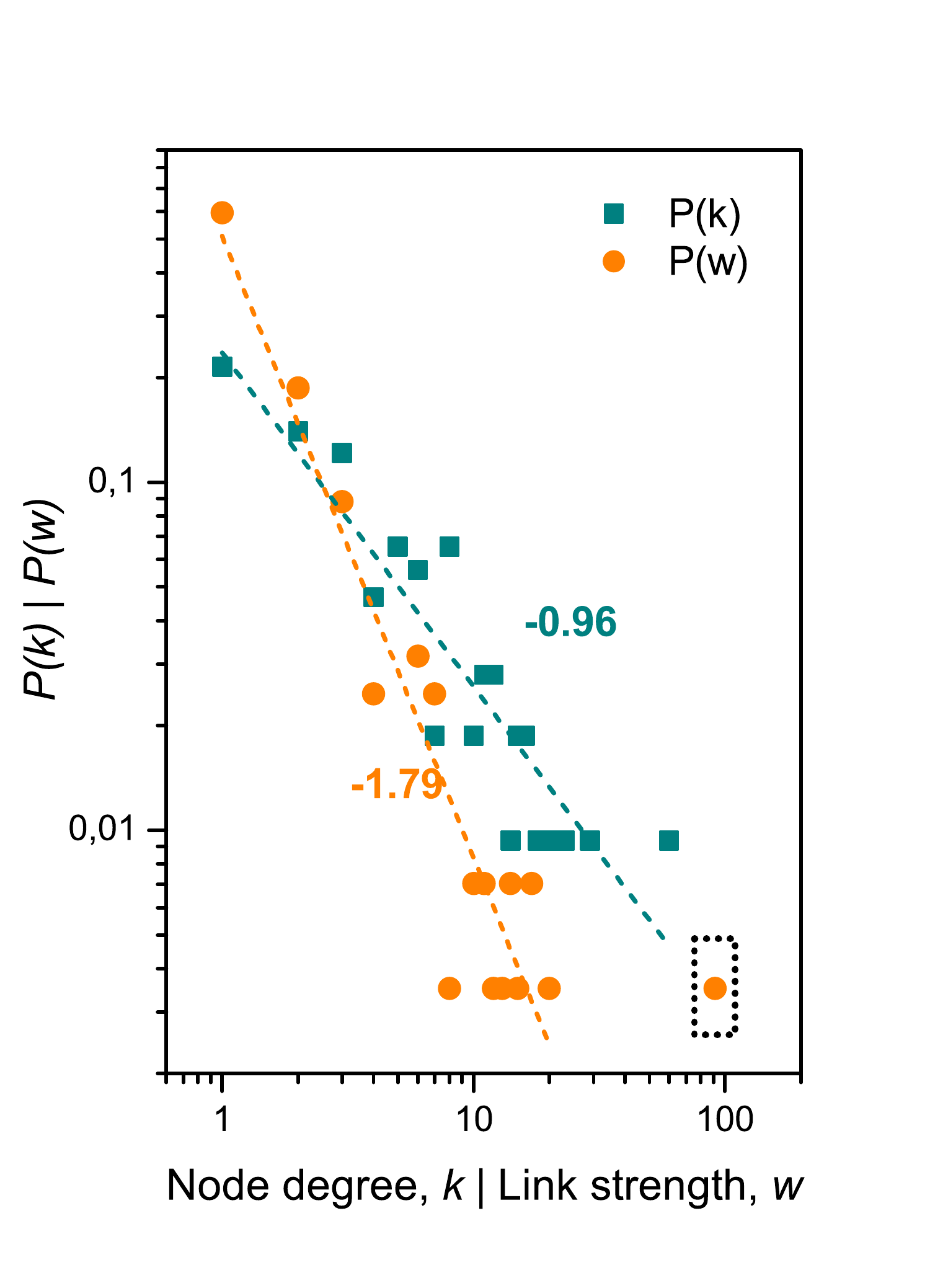}}
		\caption{Probability distributions of nodes degrees (squares) and links widths (circles) for the collaboration network at the level of countries. The numbers indicate the estimated slopes for linear approximations. The rectangle frames the outlier of $w$ (it corresponds to the link between Ukraine and Poland).}
		\label{Fig_Degr_Link_DIstr}
		\vspace{-5mm}
	\end{wrapfigure}
	The network is sparse (284 links in total), but 89.7\% of all nodes are mutually reachable being the part of the single connected component. The rest 11 nodes are isolated. Each country is connected to five others on average. Naturally, the largest hub is Ukraine (with degree $k=60$). The rest node degrees are distributed by power law with exponent close to the ``classical'' $-1$ ($-0.96$ for our data, see Figure~\ref{Fig_Degr_Link_DIstr}). This is a very typical feature of collaboration networks in general. It is interesting to note that the distribution of link widths (strengths) can be approximated by power-law as well; however, the absolute value of exponent is much larger ($-1.79$). Therefore, some natural limit for international collaboration intensity is observed. But there is a country, which is rather an outlier in this context (it is framed and not taken into account for linear approximation on Figure~\ref{Fig_Degr_Link_DIstr}) --- the strength of the link between Ukraine and Poland is much higher than one would expect. Besides Poland with 91 common papers, TOP5 of countries that are most strongly connected with Ukraine includes the USA (20), Lithuania (17), the United Kingdom (15); Germany and Czech Republic share the 5th position (14). %
	This particular list suggests typical patterns of international collaboration: governed by geographical, historical and cultural proximities that are especially typical for social science compared to the physical and life sciences, and core-periphery-like ones in the context of collaboration of more ``advanced'' and developing countries \cite{Henriksen2018,Kuld2018,Schubert2006,Choi2012}. In this respect, Poland is a very natural partner for Ukraine, being at the same time close in many ways and more developed.

	\emph{Citations analysis.} Since the majority of Ukrainian economic journals  (and 87\% of publications in our data set) are not indexed in Scopus and Web of Science, Crossref data is a unique source of citation data based on the usage of DOI numbers.
	Despite increased popularity of the Initiative for Open Citations,  deposited metadata are often incomplete \cite{Visser2021}. As long as references lists are not properly deposited and openly available, the analysis of citation data is unreliable. Therefore, only the description of our data set in respect to available citation data can be offered. 1,499 papers ($6.3\%$) are cited at least once, according to Crossref. 
	\begin{figure}[ht]
		\centerline{\includegraphics[width=0.65\textwidth]{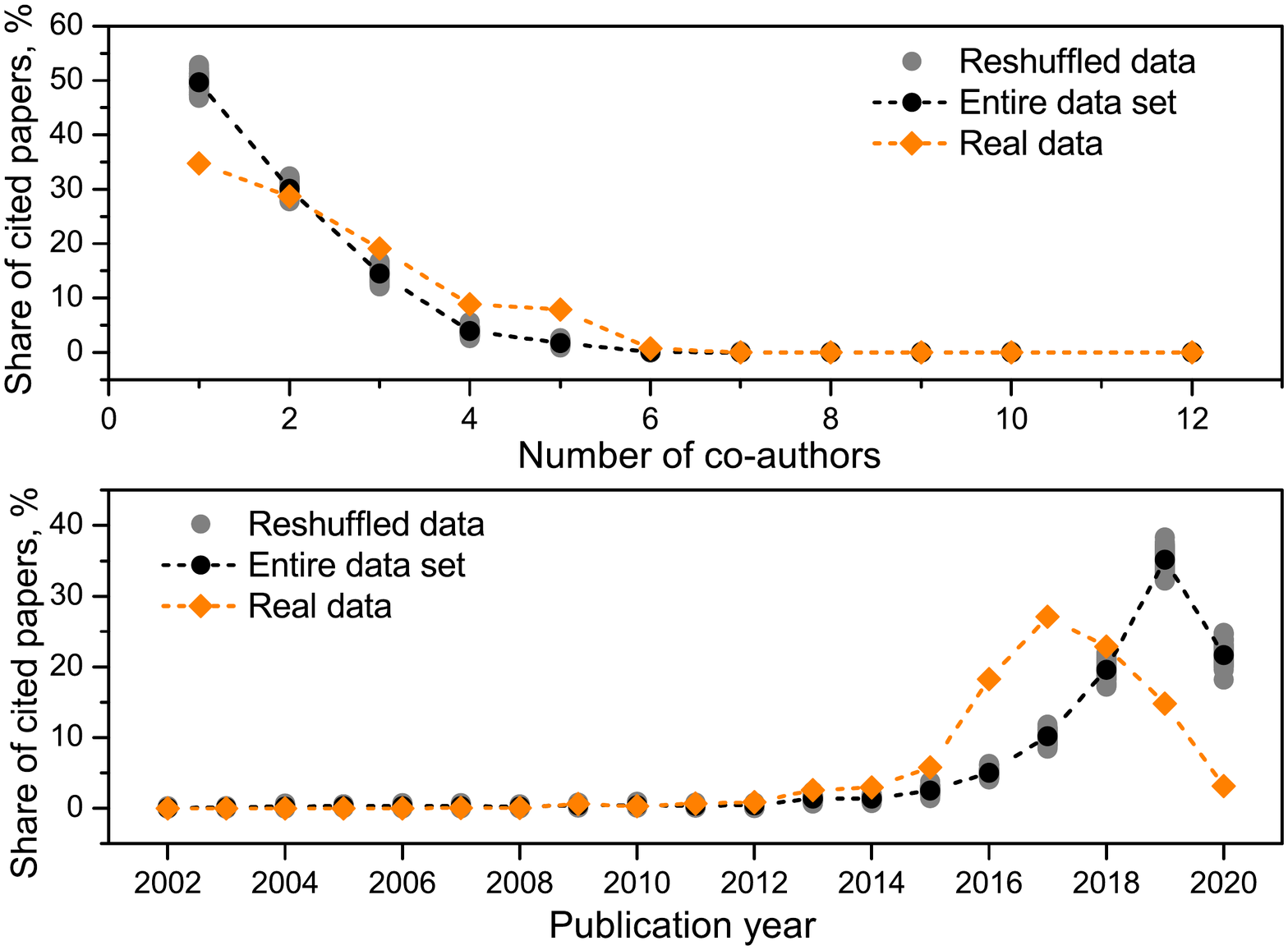}}
		\caption{Relative distributions of citation counts among papers with different numbers of co-authors (Upper panel); published in different years (Lower panel). General distributions of all papers from entire data set are shown by black circles. Distributions only for cited papers are shown by orange diamonds (real data) and grey circles (real citation counts are randomly reshuffled).}
		\label{Fig_citations}
	\end{figure}

	To see if the number of citation counts correlate with the number of co-authors, the relative distribution of cited publications is compared with the general distribution of all publications. To take into account that the subset of cited publications is comparatively small, the same distributions are also built for subsets occurred after random reassigning citation counts over entire data set (100 reshuffles were performed). Figure~\ref{Fig_citations} shows that the real distribution of cited papers differs from the distributions built for entire data set and for randomly reshuffled samples. The share of solo-publications is noticeably lower in the subset of cited publication (35\% vs. 50\%). Correspondingly, the share of cited papers with three or more co-authors is greater than expected (36\% vs. 20\%). Our results are consistent with those authors who found that collaborative papers are typically better cited, see, e.g. \cite{Kuld2018}.
	
	The geography of authors plays a role as well: 8.8\% of papers  labelled by countries are cited at least once, but this share is almost three times larger for the internationally collaborative papers (26.2\%). Of course, this is rather expected due to the fact that the majority of internationally collaborative papers is indexed in Scopus or WoS data bases. 
	
	The characteristic age of citations can be derived in a similar way. The positions of peaks, which correspond to the number of cited papers (2017) and the number of published papers (2019), allows one to suggest that for Ukrainian economic papers it typically takes two years to get cited. 
	
	\section{Conclusions}\label{Conclusions_sec}
	This work has a number of goals. First of all, this is the first large-scale quantitative analysis of the Ukrainian Economics discipline. While the majority of Ukrainian scientific journals, especially in the domain of SSH, are  not indexed in the international abstract databases such as Scopus or Web of Science, and the Ukrainian Research Information System (URIS\footnote{see https://dntb.gov.ua/en/implemented-and-perspective-projects/ukrainian-research-information-system-uris-3}) is only announced, no centralised source of Ukrainian scholarly publication metadata currently exists. However, recent changes in Ukrainain policies have led to an increase in DOI assigning to research papers and, therefore, the splash of metadata deposits to the Crossref system. This opened the door for a quantitative analysis of Ukrainian research data. Such analysis aims to better understand national scholarly disciplines.  Thus, the first goal of the paper is to provide information that can be used to support the decision making for research management. 
	
	Secondly, this is one of the first examples of a large-scale and longitudinal quantitative analysis based on the open Crossref data. In particular, the advantages of the OUCI interface for gathering Ukrainian data are discussed. It can be concluded that  Crossref metadata is a perspective source that can be used to measure the citation impact and understand trends in local scientific publication micro-system,  especially  for  non-English  publications  that are poorly  represented  in  commercial citation databases. Active support of open initiatives, such as Initiative for Open Citations, is highly encouraged in order to provide more reliable statistics for analysis of research data.
	
	Thirdly, the case study targeted at economic research is provided. It is centered on Ukraine -- an East European country with emerging economy, whose national research is understudied and influenced by peculiarities such as Cyrillic writing and non-English speaking. The obtained quantitative results are discussed and compared to others published previously and based on different data sets.
	
	Data analysis only at the level of publications is performed in this work. However, different aspects are studied: the collaborative nature of economic research, its geographic landscape, and some peculiarities of citation statistics. We found that Ukrainian economics is characterized by a comparably small share of co-authored publications, however, it demonstrates the tendency towards more collaborative output. One can speculate about the differences in the publishing behaviour of Ukrainian researchers in Economics, which depends on the targeted journals. On the one hand, ``delayed'' collaboration patterns compared to world trends are observed for publications in national journals. This gives one a reason to speculate about the dominantly  qualitative and theoretical nature of economic research at the national level. On the other hand, a higher level of international and individual cooperation is found for papers published in high-level journals indexed in Scopus or Web of Science data bases. Such papers are also typically better cited. 
	
	The geography of authors who make the contributions to Ukrainian economic journals was inspected. While part of foreign contributors collaborate with Ukrainian scholars producing co-authored outputs, the others naturally consider Ukrainian journals indexed in the international abstract data bases as a platform to efficiently disseminate their results. The proper external incentives could be  organized in order to influence both segments improving Ukrainian international collaboration structure and fortifying national scientific journals. To give an example, such national compensation or reward programs aimed at improving publishing activity exist in many countries. 
	
Finally, some future prospects for this work can be drawn. While only the most straightforward way of data analysis -- at the  publication level -- is used in this work, there is a potential to study the collected data at the level of authors. Many other aspects can be investigated: publication productivity and collaboration patterns of individual scholars, gender peculiarities, etc. However, more technical steps are required first: the peculiarities of  Ukrainian metadata, such as non-native English names and the usage of Cyrillic alphabet, complicate the process of author disambiguation. It is also planned to analyse co-authorship networks at the level of authors.


\end{document}